\newcommand{\fraz}[2]{\frac{\displaystyle #1}{\displaystyle #2}}
\newcommand{\k}{\kappa}
\newcommand{\eps}[3]{\epsilon_{#1 #2 #3}}
\newcommand{\JJ}{{\cal J}}
\newcommand{\MM}{{\cal M}}
\newcommand{\LL}{{\cal L}}
\newcommand{\PP}{{\cal P}}
\newcommand{\WW}{{\cal W}}
\newcommand{\CC}{{\cal C}}
\newcommand{\DD}{{\cal D}}
\newcommand{\g}{\gamma}

\documentstyle[12pt,a4]{article}
\begin{document}
\pagestyle{empty}

\begin{center}
\vspace*{1cm}

{\Large\bf The Quantum Deformed Dirac Equation from the \(\k\)--Poincar\'e
Algebra.}

\vspace*{3cm}

{\bf Anatol NOWICKI} \footnote{On leave of absence from Institute of Physics,
        Pedagogical University,\\ Plac S\l owia\'{n}ski~6, 65029 Zielona
        G\'ora, Poland.} ,
{\bf Emanuele SORACE and Marco TARLINI}\\
\vspace*{.5cm}
        {\small INFN, Sezione di Firenze}\\
	{\small Dipartimento di Fisica, Universit\`{a} degli Studi di
	Firenze}\\
	{\small L.go E. Fermi 2, 50125 Firenze, Italy}\\

\vspace*{3cm}

{\bf Abstract}

\end{center}

{\footnotesize
\noindent In this letter we derive a deformed Dirac equation invariant
under the \(\k\)--Poincar\'e quantum algebra. A peculiar feature is
that the square of the \(\k\)--Dirac operator is related to the second
Casimir (the \(\k\)--deformed squared Pauli--Lubanski vector). The
``spinorial'' realization of the \(\k\)--Poincar\'e is obtained by
a contraction of the coproduct of the real form of \(SO_q(3,2)\)
using the 4--dimensional representation which results
to be, up some scalar factors, the same of the undeformed algebra
in terms of the usual \(\g\)--matrices.  }

\vspace*{2.5cm}

\noindent Preprint DFF 177/12/92 Firenze, December 1992.

\pagebreak

\setcounter{page}{1}
\pagestyle{plain}

In recent papers the structure of quantum deformations of the Poincar\'e
algebra has been found and studied in detail \cite{luk1,luk2,luk3,gil1,luk4},
one of the approach used is based on the generalization of the contraction
procedure of Lie algebras to Hopf structure introduced in \cite{cele1,cele2};
in the contest
of quantum algebras the main new point is to rescale, beside the generators,
the quantum deformation parameter.
The contraction we consider in this paper is performed from
the standard real form of \(SO_q(3,2)\) using \(q\) real and defining
\(\k^{-1}=R \log q\) as the rescaled quantum parameter (\(R\) is the
de--Sitter curvature and goes to infinity in the contraction limit,
\(\k\) then has the dimension of the inverse of a length)\cite{luk4}.
This is precisely the choice proposed in \cite{cele2} to obtain the quantum
Euclidean algebra from the contraction of \(SO_q(4)\), the contracted algebra
in both the cases is a real form under an involution having the standard
properties at the level of algebra and coalgebra. In \cite{gil1}
it has been found a non linear transformation of the boost sector that
simplifies the algebra of the \(\k\)--Poincar\'e and  leads to a more
symmetric form for the coalgebra.

Let us summarize this final structure, the algebra reads:
\begin{eqnarray}
&&\ [P_i,P_j] = 0\ ,\hspace{2.26cm} [P_i,P_0]\ =\ 0\ ,\nonumber \\
&&\ [M_i,P_j] = i \eps{i}{j}{k}\; P_k\ ,\hspace{1.06cm} [M_i,P_0]\ =\ 0\ ,
\nonumber \\
&&\ [L_i,P_0] = i P_i\ ,\hspace{1.95cm} [L_i,P_j]\ =\ i\delta_{ij}\;\k\,
\sinh(P_0/\k)\ ,\\
&&\ [M_i,M_j] = i \eps{i}{j}{k}\; M_k\ ,\quad\quad [M_i,L_j]\ =\ i
\eps{i}{j}{k}\; L_k\ , \nonumber \\
&&\ [L_i,L_j] = -i \eps{i}{j}{k} ( M_k \cosh(P_0/\k) - 1/(4\k^2)\;
            P_k\;P_lM_l )\ .\nonumber
\end{eqnarray}

Where \(P_{\mu}\equiv\{P_0,P_i\}\) are the deformed energy and momenta,
the \(M_i\) are the spatial rotation
generators (they close an undeformed Hopf subalgebra), the \(L_i\) are the
deformed boost generators. The coalgebra and the antipode result:
\begin{eqnarray}
\Delta M_i &=& M_i\otimes I + I\otimes M_i\ , \quad\quad
\Delta P_0\ =\ P_0\otimes I + I\otimes P_0\ , \nonumber\\
\Delta P_i &=& P_i\otimes \exp(\frac{P_0}{2\k}) +
               \exp(-\frac{P_0}{2\k})\otimes P_i\ ,\nonumber\\
\Delta L_i &=& L_i\otimes \exp(\frac{P_0}{2\k}) +
               \exp(-\frac{P_0}{2\k})\otimes L_i + \\
            &&\fraz{1}{2\k}\; \eps{i}{j}{k}
              \left(P_j\otimes M_k \exp(\frac{P_0}{2\k}) +
                    \exp(-\frac{P_0}{2\k})M_j\otimes P_k\right)\ ,\nonumber
\end{eqnarray}
\begin{equation}
S(P_{\mu})=-P_{\mu}\ ,\quad S(M_i)=-M_i\ ,\quad
S(L_i)=-L_i+\fraz{3i}{2\k}\; P_i\ .
\end{equation}

The deformed Casimir operators are:
\begin{equation}
C_1=\left(2\k\,\sinh(\fraz{P_0}{2\k})\right)^2-P_iP_i\ \ ,\quad
C_2=\left(\cosh(\fraz{P_0}{\k})-\fraz{P_iP_i}{4\k^2}\right)\, W_0^2-W_iW_i
\end{equation}
\noindent where \cite{gil1,luk4} \(W_0=P_iM_i\) and \(
W_i=\k\,\sinh(P_0/\k)\, M_i + \eps{i}{j}{k}\; P_j L_k\ \).\\

In the present letter we want to derive the \(\k\)--deformation of the Dirac
equation: in \cite{luk4} a square root of the \(\k\)--deformed
Klein--Gordon equation (the eigenvalue equation for the invariant \(C_1\))
has been written.
The question whether that deformed Dirac equation is invariant under
``spinorial"  representations of the \(\k\)--Poincar\'e arises naturally.
To answer we build the spin--1/2 realization for the \(\k\)--Poincar\'e algebra
and we find the invariant \(\k\)--Dirac operator: it
differs from that written in \cite{luk4} which results to be not
invariant, both the operators in the classical
limit \(\k \rightarrow \infty \) produce the usual Dirac equation.

Let us consider firstly the usual Poincar\'e symmetry and  how to
build the ``spinorial" representations. In the tensor product
\(SO(3,2)\otimes SO(3,2)\) we define the operators:
\begin{equation}
\JJ_{AB}=M_{AB}\otimes I + I\otimes\Sigma_{AB}\ ,\quad A,B=0,\cdots ,4
\label{prim}
\end{equation}
\noindent where \(M_{AB}\) are a ``spinless'' realization and
\(\Sigma_{AB}\) are generators of a finite dimensional
representation of \(SO(3,2)\). The Lie structure induces the \(\JJ_{AB}\)
to close a \(SO(3,2)\) algebra. Defining \(P_{\mu}=R^{-1} M_{0\mu}\ ,
\ \mu=1,\cdots ,4\) and performing the contraction \(R\rightarrow \infty\)
the ``spinorial" representations of the Poincar\'e algebra are obtained.
The Casimir \(\JJ_{AB} \JJ^{AB}\) produces in the limit, beside the
Klein--Gordon operator \(P_{\mu}P^{\mu}\) and the finite
representation invariant \(\Sigma_{AB}\Sigma^{AB}\),
the mixed invariant \(P_{\mu}\Sigma^{0 \mu}\) \cite{cele3}.
The lowest dimension \(SO(3,2)\) representation is four by four and the
\(\Sigma^{0 \mu}\) are proportional to the usual \(\g^{\mu}\): the
Dirac equation is then written from the contracted invariant
\(P_{\mu}\Sigma^{0 \mu}\) (for higher spin this invariant gives the
Bhabha wave equations).

The same procedure can be followed in the quantum case, of course we
must substitute the structure defined in (\ref{prim}) by that derived
from the coproduct of \(SO_q(3,2)\), in the second factor
of the tensor product we substitute the 4--dimensional representation
of \(SO_q(3,2)\). It is not surprising that this representation is, up
to some factors depending on \(q\), the representation of the usual
Dirac \(\g\)--matrices. Indeed we find, following the notation of
\cite{luk4}, the 4--dimensional realization of \(SO_q(3,2)\) in the
Cartan--Weyl basis given by:
\pagebreak
\begin{eqnarray}
e_{\pm 1} &=& \pm\, 1/2\;f(\lambda) (\g_1\pm i\g_2)\g_3\ ,\nonumber\\
e_{\pm 2} &=& \ \ 1/4\; (\g_1\mp i\g_2)(\g_0\pm\g_5)\ ,\nonumber\\
e_{\pm 3} &=& -1/4\; f(\lambda) ((1+q^{\pm 1})+i(1-q^{\pm 1})\g_1\g_2)
              (\g_0\pm \g_5)\g_3\ ,\nonumber\\
e_{\pm 4} &=& -1/4\; f(\lambda)^2(1+q^{\pm 1})(\g_1\pm i\g_2)
              (\g_0\pm \g_5)\ ,\label{gamma}\\
h_1 &=& i/2\; \g_1\g_2\ ,\nonumber\\
h_2 &=& (i\, \g_3-1)h_1\ ,\nonumber\\
h_3 &=& i\, \g_3 h_1\ ,\nonumber\\
h_4 &=& (i\, \g_3+1) h_1\ ,\nonumber
\end{eqnarray}
\noindent where
\(f(\lambda)=\left(\fraz{\sin(\lambda/2)}{\sinh(\lambda)}\right)^2\; ,
\ q=\exp(\lambda)\) and \(\{\g_{\mu},\g_{\nu}\}=2g_{\mu \nu}\ \), with
\(g_{\mu \nu}={\rm diag}(1,-1,-1,-1)\ \), \(\g_5=\g_0\g_1\g_2\g_3\).

We perform the contraction limit \(R\rightarrow \infty\ \), with \(\log q
= (\k R)^{-1}\), on the coproduct of the \(SO_q(3,2)\) generators by
substituting in the second tensor factor the expressions (\ref{gamma})
in terms of the \(\g\)--matrices and defining the standard rescaling
\(M_{0\mu}=RP_{\mu}\) in the first one.
Denoting in calligraphic the global generators, in
capital italic the generators of the orbital ``spinless" part (first
tensor factor) and in small italic the ``spinorial" operators, we get:
\begin{eqnarray}
\PP_i &=& P_i\ ,\quad\quad \PP_0\ =\ P_0\ ,\nonumber\\
\MM_i &=& M_i + m_i\ ,\nonumber\\
\LL_3 &=& L_3 +  \exp(-\frac{P_0}{2\k})\; l_3\ ,\\
\LL_+ &=& L_+ + \exp(-\frac{P_0}{2\k})\; l_+ + \frac{1}{2\k}\, m_3\;
P_-\ ,\nonumber\\
\LL_- &=& L_- + \exp(-\frac{P_0}{2\k})\; l_- + \frac{1}{2\k}\, m_3\;
P_+\ ,\nonumber
\end{eqnarray}
\noindent where \(m_i=i/4\; \eps{i}{j}{k}\; \g_j\g_k\) and
\( l_i=1/2\; \g_5\g_i\ \), (\(l_{\pm}=l_1\pm i l_2\)) close an usual
\(SO(3,1)\) algebra.\\

After the change of basis defined in \cite{luk4} the final form of the
global boost generators is:
\begin{equation}
\LL_i = L_i + \exp(-\frac{P_0}{2\k})\; l_i -
        \fraz{1}{2\k}\; \eps{i}{j}{k}\; m_j\, P_k\ .
\end{equation}
These \(\LL_i\) together with the \(\MM_i\) and the \(\PP_i\ , \PP_0\)
close the algebra (1); in the limit \(\k\rightarrow \infty\) we
recover the usual ``spinorial" representation of the Poincar\'e
algebra. We notice that in the coproduct of
\(SO_q(3,2)\) the structure of the two factors of the tensor
product is not symmetric in the interchange, then the operators obtained
by substituting the 4--dimensional representation in the first tensor factor
or in the second one are different. After the limit and
the transformation of basis the formulas obtained in the two ways are related
by the change \(P_i \leftrightarrow - P_i\ , P_0 \leftrightarrow -P_0\),
this is an automorphism of the algebra (1) so that the choice is arbitrary.

We assume, as in the classical Poincar\'e case, a ``spinless''
realization for \(P_0\, ,\; P_i\, ,\; M_i\) and \(L_i\ \); we define
\begin{eqnarray}
P_0 &=& p_0\ , \hspace{2.5cm} P_i\ = \ p_i\ , \nonumber\\
M_i &=& -i\; \eps{i}{j}{k}\; p_j\, \fraz{\partial}{\partial p_k}\ ,\\
L_i &=& i\,\left( \k\,\sinh(p_0/\k)\, \fraz{\partial}{\partial p_i}\ +
p_i\, \fraz{\partial}{\partial p_0}\right)\ .\nonumber
\end{eqnarray}

They close the algebra (1) with \(W_0=W_i=0\) and therefore \(C_2=0\).

The Casimir \(\CC_1\) written in terms of the global generators does not
change from \(C_1\), the second Casimir \(\CC_2\) is obtained by
substituting for
\(W_0\) and \(W_i\) in \(C_2\) the operators
\begin{eqnarray*}
\WW_0 &=& \PP_i\, \MM_i\ =\ m_i\, P_i\ ,\\
\WW_i &=& \k\,\sinh(\PP_0/\k)\, \MM_i +
\eps{i}{j}{k}\; \PP_j \LL_k\ =\ m_i\; \k\,\sinh(P_0/\k) -
\eps{i}{j}{k}\; \tilde{l}_j P_k\ ,
\end{eqnarray*}
\noindent where \(\tilde{l}_i=\exp(-P_0/2\k)\;l_i -
1/(2\k)\;\eps{i}{j}{k}\;m_j\,P_k\ \).\\

Using the \(\g\)--matrices algebra one gets:
\begin{equation}
\CC_2=-\, C_1\left(1+\fraz{C_1}{4\k^2}\right)\; m_i\,m_i=
           -\,\frac{3}{4}\,C_1\left(1+\fraz{C_1}{4\k^2}\right)\ ,
\end{equation}
in agreement with \cite{gil1} and with the classical limit \(\k \rightarrow
\infty\).

The final result of this letter is to write a \(\k\)--deformed Dirac operator
invariant under the global generators \(\PP_0\; ,\ \PP_i\; ,\ \MM_i\;
,\ \LL_i\ \). From the commutation relations (1) it is not difficult to
show that the operator:
\begin{equation}
\DD=-\;\exp(-\frac{P_0}{2\k})\,\g_i\,P_i + i\,\g_5\,\k\,\sinh(P_0/\k)
- \fraz{i}{2\k} \g_5\, P_i\,P_i\ ,
\end{equation}
\noindent verifies \([\DD,\LL_i]=0\), and trivially \([\DD,\MM_i]=
[\DD,\PP_i]=[\DD,\PP_0]=0\). Defining \(\g_4=i\g_5\) , in the \(k
\rightarrow \infty\) limit we recover the Dirac operator.

The square of \(\DD\) results to be
\begin{equation}
\DD^2 = C_1\left(1+\fraz{C_1}{4\k^2}\right)
\end{equation}
\noindent then the second invariant takes the form:
\begin{equation}
\CC_2=-\;\frac{3}{4}\;\DD^2\ .
\end{equation}

The operator \(\DD\) is therefore the square--root of the second
Casimir \(\CC_2\) and not of \(\CC_1\), this appears as a peculiar
feature of the \(\k\)--Poincar\'e which was not {\it a priori} easy to
foresee. However we remark that a given value for \(\CC_2\) induces one
and only one positive value for \(\CC_1\).

The \(\k\)--Dirac equation is then:
\begin{equation}
\DD\ \psi = m \left(1+\frac{m^2}{4\k^2}\right)^{1/2}\ \psi\ ,
\label{dirac}
\end{equation}
\noindent where \(m^2=4\k^2\;\sinh^2(\fraz{p_0}{2\k})-p_i\, p_i\;\).\\

Using the expression for the Casimir \(\CC_1\) we can express the
operator \(D\) in the following form
\begin{equation}
\DD=\exp(-\frac{P_0}{2\k})\left(-\,\g_i\,P_i + i\,\g_5\,2\k\,
\sinh(\frac{P_0}{2\k})\right)+ i\,\g_5\,\fraz{\CC_1}{2\k}\ .
\end{equation}
Thus for the massless case the \(\k\)--Dirac equation coincides with
the one  proposed in \cite{luk4}.

Assuming to be on ``mass shell'', \(\CC_1=m^2\), the \(\k\)--Dirac
wave equation is linear with respect to the space derivatives while
it is a finite difference equation in time with the fundamental shift
given by \(i/(2\k)\).

In a recent paper \cite{gil2} a deformed Dirac equation is proposed
using the theory of the covariant wave functions (a generalization of
the Weinberg's approach), however it appears that this equation is
invariant under the \(\k\)--Poincar\'e algebra only for the solutions
on mass shell; we notice that the quantum deformed Dirac equation
(\ref{dirac}) is algebraically invariant also off mass shell.

We want to stress moreover that the choice of the standard real form
of the \(\k\)--Poincar\'e algebra given in \cite{luk4} allows
to represent both sector of the Hopf algebra in the Hilbert space with
a positive definite metric.

The next step in our approach is to consider the coupling of Eqn. (\ref{dirac})
with the electromagnetic field in the framework of \(\k\)--deformed
Poincar\'e symmetry, this line of research is under investigation.

\end{document}